\pdfoutput=1
\documentclass[sigconf]{acmart}
\AtBeginDocument{%
  }

\copyrightyear{2026}
\acmYear{2026}
\setcopyright{cc}
\setcctype{by}
\acmConference[WWW '26]{Proceedings of the ACM Web Conference 2026}{April 13--17, 2026}{Dubai, United Arab Emirates}
\acmBooktitle{Proceedings of the ACM Web Conference 2026 (WWW '26), April 13--17, 2026, Dubai, United Arab Emirates}
\acmPrice{}
\acmDOI{10.1145/3774904.3792948}
\acmISBN{979-8-4007-2307-0/2026/04}
\usepackage{enumitem}

\usepackage{caption}
\usepackage{subcaption}
\usepackage{natbib}
\usepackage{graphicx}
\usepackage{booktabs}
\usepackage{multirow}

\newlength{\imgsize}
\newcommand{\img}[2]{%
\includegraphics[height=\imgsize,
    width=\imgsize,keepaspectratio]{images/category_examples/#1/#2}
}

\acmSubmissionID{srp726}

\begin{document}

\title{Deepfakes in the 2025 Canadian Election: Prevalence, Partisanship, and Platform Dynamics}

\author{Victor Livernoche}
\orcid{0000-0002-3819-730X}
\affiliation{
  \institution{McGill~University,~Mila}
  \city{Montréal}
  \country{Canada}
}
\email{victor.livernoche@mail.mcgill.ca}

\author{Andreea Musulan}
\orcid{0009-0002-2964-6414}
\affiliation{
  \institution{Université~de~Montréal,~Mila}
    \city{Montréal}
  \country{Canada}
}
\email{andreea.musulan@gmail.com}

\author{Zachary Yang}
\orcid{0009-0009-6765-2293}
\affiliation{
  \institution{McGill~University,~Mila}
  \city{Montréal}
  \country{Canada}
}
\email{zachary.yang@mail.mcgill.ca}

\author{Jean-François Godbout}
\orcid{0009-0008-1886-4890}
\affiliation{
  \institution{Université~de~Montréal,~Mila}
    \city{Montréal}
  \country{Canada}
}
\email{jean-francois.godbout@umontreal.ca}

\author{Reihaneh Rabbany}
\orcid{0000-0003-2348-0353}
\affiliation{
  \institution{McGill~University,~Mila}
  \city{Montréal}
  \country{Canada}
}
\email{reihaneh.rabbany@mcgill.ca}

\renewcommand{\shortauthors}{Victor Livernoche, Andreea Musulan, Zachary Yang, Jean-François Godbout, \& Reihaneh Rabbany}

\begin{abstract}
Concerns about AI-generated political content are growing, yet there is limited empirical evidence on how deepfakes appear and circulate across social platforms during major events in democratic countries. We analyze the 2025 Canadian federal election across X, Bluesky, and Reddit using a high-accuracy detector trained on diverse modern generative models, covering \textbf{187,778 posts}. We find that \textbf{5.9\%} of election-related images were deepfakes. Right-leaning accounts shared them more often (\textbf{9.2\%} of images flagged) than left-leaning users (\textbf{3.9\%}), with flagged content more frequently defamatory or conspiratorial. Yet, most detected deepfakes were benign or non-political, and harmful ones drew little attention, accounting for only \textbf{0.1\%} of all views on X. Overall, deepfakes were present in the election conversation, but their reach was modest, and realistic fabricated images, although less common, drew higher engagement, highlighting growing concerns about their misuses.
\vspace{-0.5em}
\end{abstract}

\ccsdesc[500]{Human-centered computing~Social content sharing}
\ccsdesc[300]{Networks~Online social networks\vspace{-0.5em}}
\keywords{Deepfake Detection; AI-Generated Content; Social Media}

\maketitle
\vspace{-0.5em}
\section{Introduction}
\label{sec:intro}

\begin{figure*}[t!]
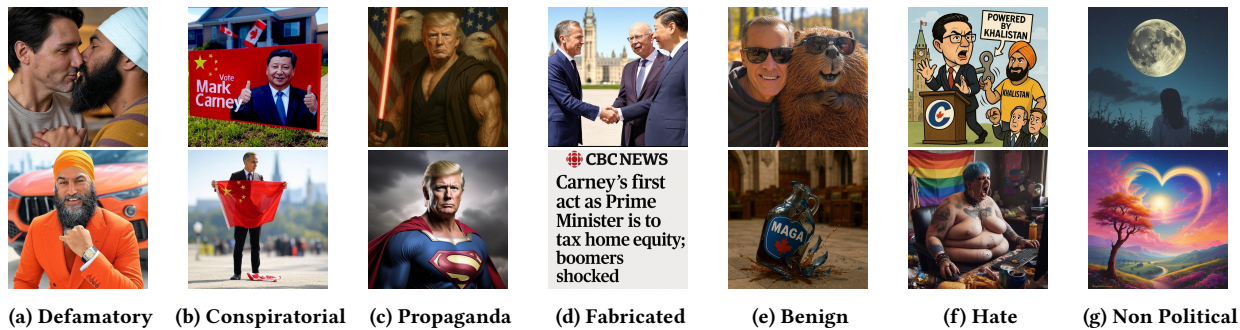

    \begin{subfigure}[b]{0.13\textwidth}
    \centering
    \img{A_Defamatory}{1.pdf}
    \caption{Defamatory}
    \end{subfigure}
    \begin{subfigure}[b]{0.13\textwidth}
    \centering
    \img{B_Conspiratorial}{1.pdf}
    \caption{Conspiratorial}
    \end{subfigure}
    \begin{subfigure}[b]{0.13\textwidth}
    \centering
    \img{C_Propaganda}{1.pdf}
    \caption{Propaganda}
    \end{subfigure}
    \begin{subfigure}[b]{0.13\textwidth}
    \centering
    \img{D_Fabricated}{1.pdf}
    \caption{Fabricated}
    \end{subfigure}
    \begin{subfigure}[b]{0.13\textwidth}
    \centering
    \img{E_Benign}{1.pdf}
    \caption{Benign}
    \end{subfigure}
    \begin{subfigure}[b]{0.13\textwidth}
    \centering
    \img{F_Hate}{1.pdf}
    \caption{Hate}
    \end{subfigure}
    \begin{subfigure}[b]{0.13\textwidth}
    \centering
    \img{G_Non_Political}{1.pdf}
    \caption{Non Political}
    \end{subfigure}
    \vspace{-10pt}
    \caption{Examples of the seven \textit{communicative intent categories} used in our analysis, illustrating the range of political and non-political uses of AI-generated imagery, from defamatory and conspiratorial content to benign and artistic creations.}\vspace{-10pt}
    \label{fig:category_examples}
\end{figure*}

Recent advances in image and video generation have made it trivial for ordinary users to create photorealistic yet entirely fabricated content. As these tools become increasingly accessible, their outputs are being shared widely across social media, often without disclosure of their artificial origin.

This rapid diffusion of generative technology has profound implications for public discourse. Deepfakes, or AI-generated media, can erode public trust, amplify misinformation, and be strategically weaponized to ridicule, glorify, or manipulate political actors. The proliferation of such media poses a pressing question: \emph{how are deepfakes being used in political communication, and to what extent do they shape the information ecosystem during democratic moments?}

To address this question, we analyze one week of social media activity surrounding the 2025 Canadian federal election (April 24--29) across X, Bluesky, and Reddit. We evaluate each image post with a state-of-the-art deepfake detector and annotate posts with communicative intent \footnote{We use “communicative intent” as an operational label of the message implied by the post’s observable content (image+text+context), not the author’s true motivation.} and author political leaning, enabling us to quantify both the prevalence and the purpose of deepfakes within online political discourse.
Specifically, our study aims to answer the following research questions:

\begin{itemize}[noitemsep, leftmargin=*, topsep=0pt, parsep=0pt, partopsep=0pt]
    \item What is the prevalence of deepfakes on major social media platforms during the election period?
    \item What \emph{perceived} communicative intents characterize the deepfakes that circulate in this context?
    \item Does political leaning affect the likelihood or manner in which deepfakes are shared?
    \item What is the impact of deepfake posts in terms of engagement and visibility?
\end{itemize}

Across the three platforms, we find that synthetic imagery is present but not dominant: about \textbf{5.9\%} of all shared images are deepfakes, with \textit{X} showing the highest rate. Most cases are benign or non-political, although politically targeted content appears more often among right-leaning accounts. Platform ecosystems differ, with \textit{X} hosting more harmful material while Bluesky and Reddit are largely non-political. Deepfakes also show limited reach. They receive fewer views when compared to real images. Overall, deepfakes appear in election discourse, but their visibility and impact during this period were modest.
\vspace{-1em}
\section{Related work}
\label{sec:related}
Recent work has begun to quantify how AI-generated visuals circulate in political settings. \citet{chen2025synthetic} document the prevalence and spread of political deepfakes on \textit{X} during the 2024 U.S. election, while \citet{chang2024generative} show how generative tools reshape meme creation and their use in political discourses. In parallel, \citet{drolsbach2025characterizing} analyze AI-generated misinformation at scale, emphasizing the need to study diffusion and engagement rather than detection alone.

Outside explicit political contexts, synthetic imagery has become commonplace in online spaces optimized for engagement. \citet{diresta2024spammersscammersleverageaigenerated} describe Facebook pages that mass-produce attention-grabbing AI images for growth and monetization, while work on creative communities (e.g., art subreddits) traces evolving norms and moderation responses to generative media \citep{matatov2025examiningprevalencedynamicsaigenerated}. A related line focuses on identity deception via AI-generated profile images: large audits report extensive use of GAN-based faces among inauthentic Twitter/X accounts~\citep{ricker2024ai}, and broader ecosystem analyses map behavioral markers and risks~\citep{Yang_2024}. These address account-level authenticity, whereas our interest is post-level content and its communicative role.
Compared to prior work, we provide the first synchronized cross-platform measurement of election-related deepfakes across \textbf{X, Bluesky, and Reddit}, using our modern detector trained on a diverse mix of proprietary and open-source generators. Prior studies often rely on community notes \citep{drolsbach2025}, heuristic labels \citep{diresta2024spammersscammersleverageaigenerated}, or prompt-based systems (e.g., GPT-4o) \citep{drolsbach2025characterizing, chen2025synthetic, chang2024generative}; our detector offers stronger accuracy and generalization. Beyond detection, we add \textbf{vision–language intent classification}, \textbf{author-leaning inference}, and \textbf{engagement analysis}, allowing us to also characterize what deepfakes aim to communicate and how far they spread.
\vspace{-1em}
\section{Methodology}
\label{sec:methodology}
We analyze images shared on three major social media platforms, \textbf{X}, \textbf{Bluesky}, and \textbf{Reddit}, during the week surrounding the 2025 Canadian federal election (April 24 to April 29). The election occurred on April 28, so the dataset captures the immediate lead-up and the day following the vote. Posts were collected using official APIs and platform-specific scraping tools under research compliance guidelines.
For X, we relied on a subscription to the \texttt{Pro} API to stream election-related posts filtered by a comprehensive list of keywords (e.g., \textit{“carney,” “poilievre,” “freeland,” “\#cdnpoli,” “\#canadapolitics”}). Only posts containing an image were retained; we excluded retweets and posts from users with fewer than 100 followers to limit spam. Similar election-related queries were applied to Bluesky and Reddit using their respective public APIs, again filtering for posts that included attached media (for Reddit, using a set of subreddits).

\vspace{-10pt}
\subsection{Deepfake Detection}
\label{sec:deepfake_detection}
We detect AI-generated images using a model built primarily on our \texttt{OpenFake} dataset \citep{livernoche2025openfakeopendatasetplatform}, which forms the foundation of the detector and is specifically designed for real-world deepfake detection in political misinformation contexts. \texttt{OpenFake} pairs over \textbf{three million real images} with approximately \textbf{one million synthetic counterparts} generated by state-of-the-art text-to-image systems spanning both open and proprietary architectures (e.g., \texttt{SDXL}, \texttt{Imagen}, \texttt{DALL·E}, \texttt{Flux}, \texttt{Ideogram}). The dataset emphasizes political and news-related content, derived from \texttt{LAION-400M} captions filtered with a vision–language model (\texttt{Qwen2.5-VL}) to retain human faces and salient sociopolitical scenes. Synthetic counterparts were generated using the same captions, ensuring tight semantic alignment between real and fake distributions.

We trained a \texttt{ConvNeXt-V2-Base} backbone for binary classification on a combined corpus exceeding \textbf{five million images}, integrating \texttt{OpenFake} with \texttt{DF40} \citep{yan2024df40} and \texttt{GenImage} \citep{zhu2023genimage} to broaden coverage of generation techniques. Training employed a mixed real–synthetic sampling strategy with extensive image augmentations (JPEG compression, color jitter, random crops) to simulate real-world degradations and manipulations.

To validate the model’s effectiveness, we used the \textit{in-the-wild} evaluation split from \cite{livernoche2025openfakeopendatasetplatform}, comprising 1,057 real and 163 deepfake-labeled images manually verified through reverse image search and metadata inspection. This evaluation subset is drawn directly from the same social media data analyzed in the present study, ensuring a consistent domain match between model validation and downstream analysis. The \texttt{ConvNeXt-V2} detector achieves a f1-score of 0.852 on this evaluation set, with 99.2\% accuracy on real images  (score $\leq 0.5$) and 77.9\% accuracy on fake images  (score $> 0.5$). This ensures we are not overestimating the reported statistics on detected deepfakes.

\subsection{Intent Classification}
\label{sec:intent_classification}

For each image flagged as a likely deepfake, we classify the post’s perceived communicative intent using a vision–language model (VLM), Qwen3-VL-32B-Instruct, prompted with the image and its accompanying text. The prompt asks the VLM for a brief rationale, a category name, and a one-letter code. We identified seven intent categories: \textit{Defamatory} (mocking or attacking individuals or groups), \textit{Conspiratorial} (suggesting hidden control or corruption), \textit{Propaganda} (promoting leaders, parties, or ideologies), \textit{Fabricated} (mimicking realistic or news imagery), \textit{Benign} (metaphoric/symbolic or satirical political content), \textit{Hate} (targeting identity groups), and \textit{Non-political} (artistic or unrelated to politics). Examples appear in Figure~\ref{fig:category_examples}, drawn directly from posts in our collected dataset.

\subsection{Political Leaning Classification}
\label{sec:leaning_classification}

We estimate the political leaning of users using a separate large language model (Llama 3.3 70B Instruct) prompted on their recent tweet history. For each unique author, we build a chronological window of their five most recent tweets, plus a system message describing the Canadian party landscape (for example, Poilievre and the Conservatives as right-leaning, Trudeau and Carney as left-leaning Liberals, etc.), to the model. The model is instructed to provide an explanation before answering. Manual inspection of 200 authors showed accurate labeling, including appropriate use of “unknown.” The main limitation is that a five-tweet window often lacks clear political cues, leading to many “unknown” labels.

\section{Results}
\label{sec:results}
\begin{table}[b]
\centering
\small
\begin{tabular}{lcccc}
\toprule
\textbf{Platform} & \textbf{Leaning} & \textbf{Total posts} & \textbf{Deepfakes} & \textbf{Share (\%)} \\
\midrule
\multirow{3}{*}{X}        & Left    & 27{,}306 & 1{,}761 & 6.45 \\
                          & Right   & 51{,}764 & 5{,}022 & 9.70 \\
                          & Unknown & 42{,}098 & 2{,}733 & 6.49 \\
\midrule
\multirow{3}{*}{Bluesky}  & Left    & 36{,}903 &   720   & 1.95 \\
                          & Right   &  3{,}096 &    53   & 1.71 \\
                          & Unknown & 25{,}171 &   668   & 2.65 \\
\midrule
\multirow{3}{*}{Reddit}   & Left    &     62   &     7   & 11.29 \\
                          & Right   &     41   &     0   & 0.00 \\
                          & Unknown &   1{,}337 &    41   & 3.07 \\
\bottomrule
\end{tabular}
\caption{Deepfake prevalence by platform and inferred political leaning. Percentages indicate the fraction of deepfakes among all collected posts within each group.}
\label{tab:deepfake_prevalence}
\vspace{-2.5em}
\end{table}

We first measured the overall prevalence of synthetic imagery during the election period. Across all platforms, \textbf{5.9\%} of shared images were detected as deepfakes, with the highest rate on \textit{X} (\textbf{7.9\%}), followed by Reddit (\textbf{3.3\%}) and Bluesky (\textbf{2.2\%}). While deepfakes remained a minority of total visual content (see Table \ref{tab:deepfake_prevalence}), their presence was nontrivial given the political context and short observation window. We note that the overall deepfake prevalence might be slightly higher than what we report. As discussed in Section~\ref{sec:deepfake_detection}, the detector does not detect all synthetic images, hence our reported prevalence underestimates the true rate, although this is unlikely to affect the overall comparative patterns we report, e.g., across platforms and political leanings.

\begin{figure}[]
    \centering
    \includegraphics[width=1.0\linewidth, trim=0 5 0 0, clip]{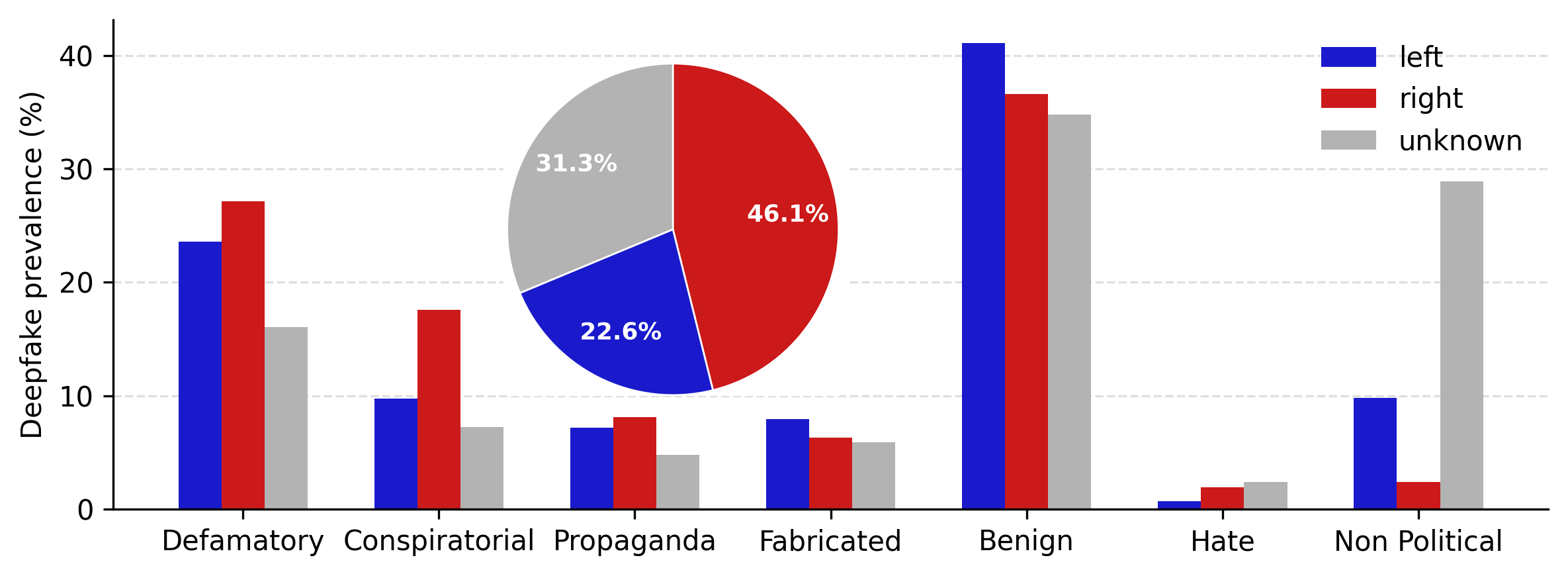}
      \vspace{-20pt}
    \caption{Distribution of deepfake intents by political leaning. Bars show the prevalence of each intent category among deepfakes posted by left-, right-, and unknown-leaning accounts, revealing distinct narrative patterns—right-leaning accounts emphasize defamatory and conspiratorial content more heavily. The pie chart summarizes the overall share of deepfakes contributed by each leaning group.}
    \vspace{-5pt}
    \label{fig:deepfake_by_leaning}
\end{figure}

\begin{figure}[]
    \centering
    \includegraphics[width=1.0\linewidth, trim=0 5 0 0, clip]{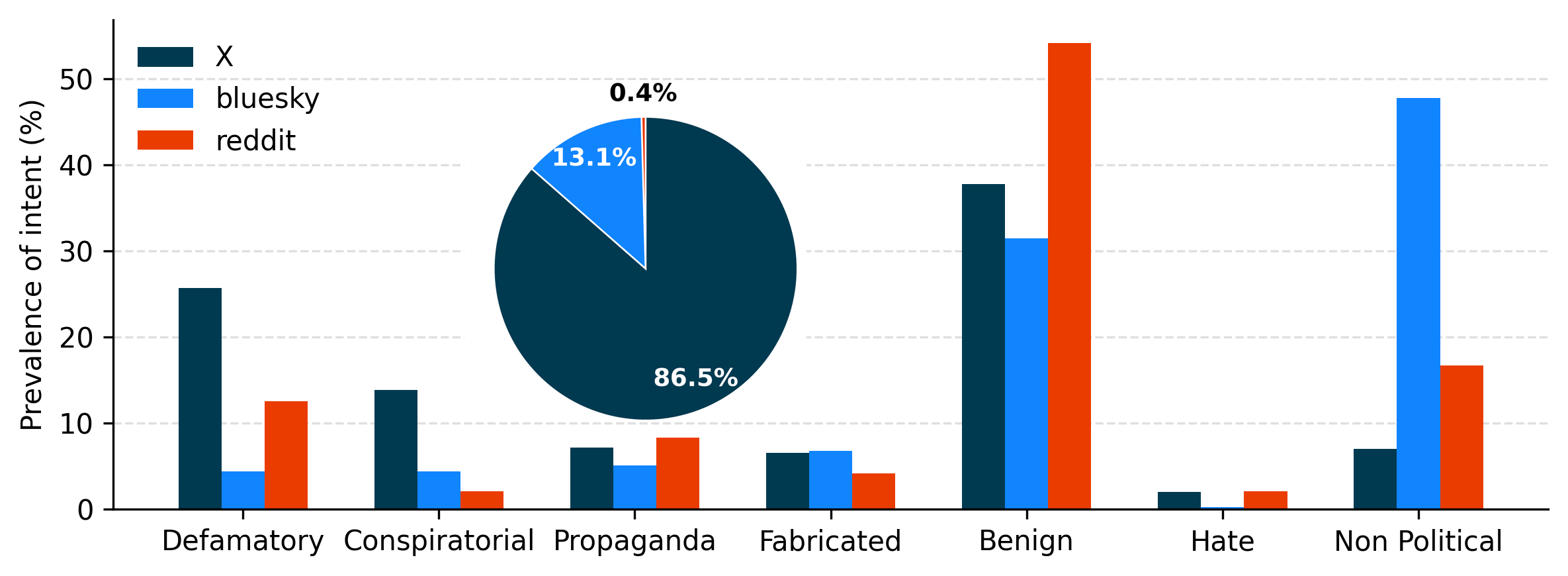 }
     \vspace{-20pt}
    \caption{Prevalence of deepfake intents across platforms. Bars show the percentage of detected deepfakes assigned to each intent category on X, Bluesky, and Reddit, highlighting platform-specific differences in the types of synthetic content circulating during the election period. While this histogram is normalized per platform,  the pie chart inset summarizes the overall share of deepfakes across platforms.}
    \vspace{-10pt}
    \label{fig:deepfake_by_platform}
\end{figure}

Deepfake prevalence varies by inferred political orientation.
As shown in Figure~\ref{fig:deepfake_by_leaning}, right-leaning users had the highest deepfake rate, with \textbf{9.2\%} of their posts flagged as synthetic, compared to \textbf{3.9\%} for left-leaning users and \textbf{5.0\%} for neutral or unidentified accounts. Intent distributions were broadly similar across leanings, though right-leaning posts skewed more conspiratorial, while left-leaning posts contained more non-political content. A detailed breakdown of posting activity and deepfake shares confirms that \textit{X} contains a larger proportion of right-leaning accounts and Bluesky contains a larger proportion of left-leaning accounts, which provides a basic sanity check that our political-leaning classifier behaves as expected. Figure~\ref{fig:deepfake_by_platform} further shows the relative frequency of deepfake intent categories across the three platforms. Most detected deepfakes fall into the Benign class, typically memes, illustrative content, or humorous edits without clear deceptive cues. Deepfakes on \textit{X} were far more often defamatory, whereas those on Bluesky were predominantly non-political or benign.

\begin{figure}[H]
\centering\includegraphics[width=1.0\linewidth, trim=5 7 2 8, clip]{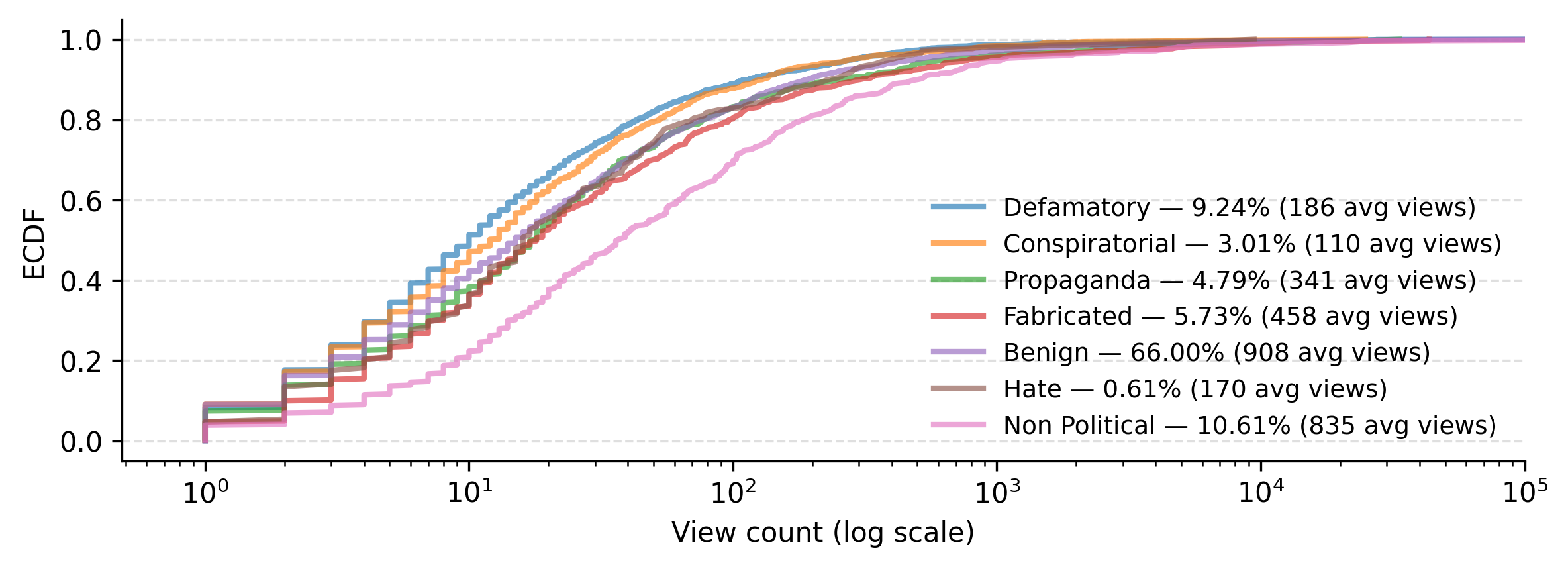}
    \vspace{-20pt}
    \caption{View counts of intents. Each curve shows the ECDF: the share of posts below a given view count. Steeper early rises indicate lower exposure. Non-political deepfakes tend to reach higher view counts, while political ones remain less viewed. The legend reports the proportion of total deepfake views and the average views per post for each category.}
    \vspace{-1em}
    \label{fig:ecdf_category}
\end{figure}
\begin{figure}[H]
\vspace{-10pt}
\centering
    \includegraphics[width=0.98\linewidth, trim=5 7 5 10, clip]{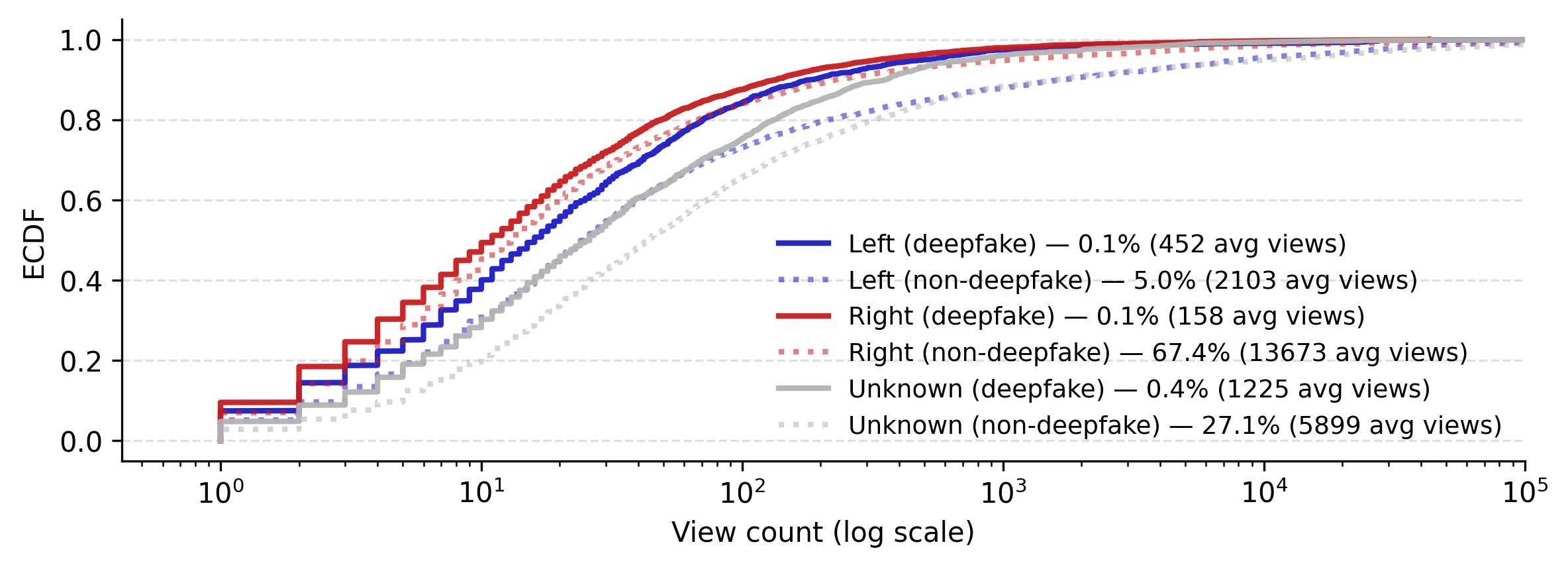}
    \vspace{-10pt}
    \caption{View counts by political leaning and deepfakes status. Within every political leaning, the deepfake curves rise earlier than their non-deepfake counterparts, showing that deepfake posts typically attract fewer views. The legend shows total view share and average views per post.
    }
    \vspace{-10pt}
    \label{fig:ecdf_leaning}
\end{figure}

To evaluate reach, we compared view counts on \textit{X} six months after posting. The empirical cumulative distribution functions (ECDFs) in Figures~\ref{fig:ecdf_category} and~\ref{fig:ecdf_leaning} show that deepfake posts consistently received fewer views than non-deepfake posts across all political leanings. While we use views, likes and reposts yield similar results.%

Most detected deepfakes were benign or non-political (together \textbf{49.4\%} of cases), and deepfakes overall accounted for only \textbf{0.5\%} of all views on \textit{X}. Consequently, only \textbf{0.1\%} of total views corresponded to politically harmful deepfakes. Benign and non-political synthetic images achieved slightly higher median exposure, while politically charged categories, especially \textit{Defamatory} and \textit{Hate}, remained concentrated in the low-view tail with minimal amplification. An important exception is the \textit{Fabricated} category, which, although less common, received more views than most categories aside from benign and non-political content. A similar pattern held for likes, which closely followed view counts and showed equally low engagement for politically charged deepfakes. Moreover, deepfakes from accounts with unknown leaning attracted more views than those from left- or right-leaning users. Deepfake and non-deepfake posts on \textit{X} were removed (deleted after 6 months) at nearly identical rates (13.8\% vs.\ 13.0\%), indicating that deepfakes were not disproportionately targeted for removal.

\begin{figure}[H]
\centering
    \includegraphics[width=0.98\linewidth, trim=5 5 5 7, clip]{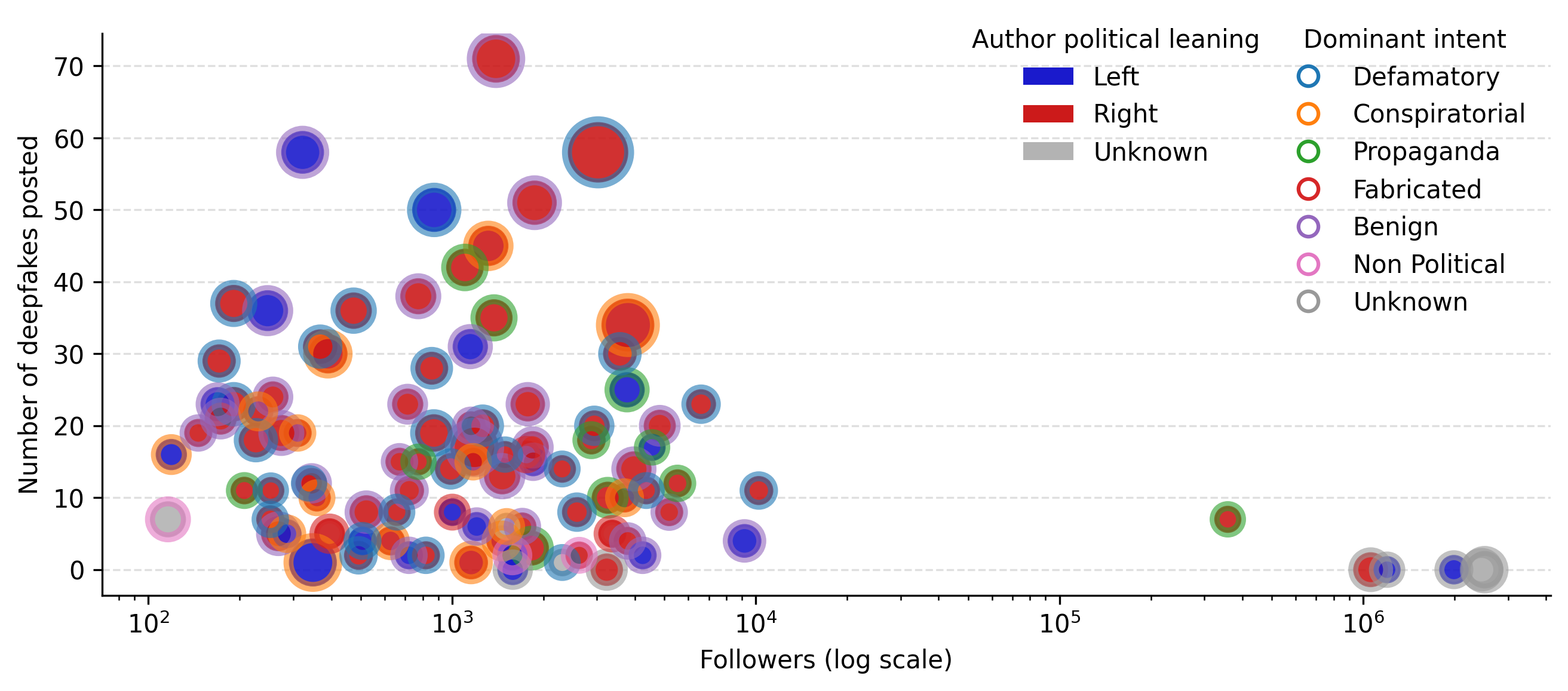}
    \vspace{-10pt}
    \caption{Number of deepfakes vs. author follower count (log scale). Each point is an author on X, colored by political leaning, sized by total posts, and ring-colored by dominant deepfake use. Most deepfake-heavy users have small followings, while high-follower accounts share few deepfakes.}
    \vspace{-10pt}
    \label{fig:author_deepfake_scatter_number}
\end{figure}

Figure~\ref{fig:author_deepfake_scatter_number} plots the number of deepfakes posted per author against follower count on X. Most prolific deepfake sharers have modest followings, indicating that the spread of deepfakes is driven by small or fringe accounts rather than influential figures. A few high-follower accounts do post deepfakes, but infrequently. The most common intent among low-volume deepfake sharers is benign, whereas several high-volume deepfake accounts predominantly share defamatory or conspiratorial content.

\section{Discussion and Conclusion}
\label{sec:discussion}
Our results suggest that synthetic images were present during the 2025 Canadian election period but did not substantially shape online attention. Deepfakes circulated across platforms, yet most reached only small audiences and were posted by low-visibility accounts. This pattern implies that synthetic content is becoming part of everyday online expression, but its current influence on large-scale political discourse remains limited.

A more important insight comes from which deepfakes gain traction. Highly realistic fabricated images, although relatively rare, received more engagement than most politically oriented categories. Because these visuals can resemble genuine news, they carry a higher risk of misleading users and triggering cascades of resharing. Even if the overall volume is modest, the plausibility of these images creates a disproportionate vulnerability in moments of political tension. Low-volume but high-credibility synthetic content can still sway perceptions, especially if deployed strategically or circulated within tightly connected communities. The threat does not come from a flood of deepfakes, but from targeted fabrications that appear credible enough to be taken as authentic evidence.

Our estimates are conservative, as the detector favors precision and therefore misses some synthetic images. Many items classified as non-political deepfakes are simple logo edits or overlays, and the small number of Reddit images reduces the statistical significance of its results. These caveats do not change the broader pattern but highlight the need for stronger and more accessible detection tools. As synthetic media becomes easier to create and harder to distinguish from reality, we will need reliable, publicly usable systems to flag deepfakes and preserve trust in democratic information.
\vspace{-0.4em}
\begin{acks}
This work was supported by the CIFAR AI Chairs Program, CSDC, IVADO, and the Canada First Research Excellence Fund. We also thank Mila for financial and computational support, and Fuxiao Gao for assistance with data collection.
\end{acks}

\vspace{-0.4em}
\bibliographystyle{ACM-Reference-Format}
\bibliography{references}

\appendix

\end{document}